# AUDIO CODEC ENHANCEMENT WITH GENERATIVE ADVERSARIAL NETWORKS

*Arijit Biswas[1] and Dai Jia[2]*

[1]Dolby Germany GmbH, Nürnberg, Germany
[2]Dolby Laboratories, Beijing, China

## ABSTRACT

Audio codecs are typically transform-domain based and efficiently code stationary audio signals, but they struggle with speech and signals containing dense transient events such as applause. Specifically, with these two classes of signals as examples, we demonstrate a technique for restoring audio from coding noise based on generative adversarial networks (GAN). A primary advantage of the proposed GAN-based coded audio enhancer is that the method operates end-to-end directly on decoded audio samples, eliminating the need to design any manually-crafted frontend. Furthermore, the enhancement approach described in this paper can improve the sound quality of low-bit rate coded audio without any modifications to the existent standard-compliant encoders. Subjective tests illustrate that the proposed enhancer improves the quality of speech and difficult to code applause excerpts significantly.

## 1. INTRODUCTION

Popular audio coding solutions are lossy [1], i.e. they reproduce audio only with reduced fidelity. Audio coding artifacts [2] are perceived due to the introduction of quantization noise. Typically, these audio coding systems are based on transform coding in the MDCT-domain. Within a frame of MDCT, audio codecs normally shape the coding noise in the frequency domain to make it least audible: exploiting the so-called simultaneous masking [1] phenomenon. Several successful audio codecs utilize frames of long durations to maximize the coding gain for stationary signals. Since the quantization noise introduced into the signal during the encoding process spreads uniformly throughout the transform block during decoding, coding noise may be most audible during low-intensity passages within an MDCT frame. For example, quantization noise in the quiet region prior to a transient event in a decoded audio signal can be perceived as pre-echoes [1],[2] at the attack portions of transient signals or as reverberation in speech signals [2]. Several coding tools were introduced in state-of-the-art audio codecs to handle speech effectively (see [3], and the references therein). While these tools are effective in modeling the series of pseudo-stationary events in speech, coding speech with very low bitrate audio codecs introduces time smearing and reverberation artifacts. In most of the state-of-the-art audio coding systems, there is a dedicated speech codec [4],[5] designed to operate on speech at very low bitrates, where it offers an advantage over the audio codec. However, there is a remaining opportunity to enhance the perceptual quality of audio codecs for a broader class of nonstationary signals and bitrates.

Applause signals are even more challenging for audio codecs. These signals can be viewed as a sound texture composed of a noise-like background (due to dense random background claps) and distinct transient clapping events in the foreground [6],[7]. Both signal components are difficult to code for a transform coder. While the noise-like component cannot be decorrelated well due to its random nature, the transient claps are expanded (rather than compacted) in the transform-domain representation. Moreover, any quantization noise introduced by the coder will be smeared over the duration of a transform block leading to pre- and post-echoes (if long blocks are used). Thus, low bitrate audio coding of an applause signal leads to smeared transients and a perceived increase in noise-like signal character. Again, readers are referred to [3], and references therein for a list of state-of-the-art applause coding tools. Recently a post-processing technique for enhancing coded applause-like signals was proposed in [6] which is based on decomposing a signal into foreground transient (claps) events and a more noise-like background part. The method was designed to restore the transient-to-noise ratio of the decoded applause signal using a static frequency-dependent restoration profile.

The restoration of audio from coding noise is a challenging problem. Intuitively one may consider coding artifact reduction and de-noising to be highly related. However, removal of coding artifacts that are highly correlated to the desired sounds appears to be more complicated than removing other noise types that are often less correlated. Furthermore, coding artifact reduction is more complex than just spectral bandwidth extension. The characteristics of coding artifacts depend on the codec and the employed coding tools, and the selected bitrate [2]. Classic codec enhancement methods [8],[9] are targeted at specific artifacts, they require very specialized knowledge about the codec and its' parameter settings, and hence are not easy to generalize. Furthermore, such backwards compatible enhancement methods do not provide significant audio quality improvement.

The advent of deep learning has opened wide opportunities to develop backwards compatible methods for coded audio restoration [10]–[14]. Convolutional neural networks (CNN) are quite popular for the purpose. Zhao et al. [11], was the first to successfully demonstrate backwards compatible quality enhancement of narrow-band and wide-band speech codecs using a convolutional autoencoder with skip-connections (operating in time- or cepstral-domain). Porov et al. [12] proposed a novel backwards compatible CNN-based (operating on complex MDCT magnitude spectrum) mp3 coded music enhancer that utilizes correlation and context information across spectral and temporal dependency. In [14] a time-frequency long short-term memory (LSTM) recurrent neural networks (RNN), equipped with two separate LSTM-RNN layers intended to model two-dimensional time-frequency information was used to enhance mp3 coded music.

We believe that to enhance coded audio by recovering missing information from compressed sources, we require a (conditioned) generative model [15]. Novel samples created by the generator are ideally suited to restore information lost due to coding (e.g. by intelligent gap filling). Recently, autoregressive generative models are being adopted for the purpose. In [10], the authors attempted to reduce artifacts in mp3 using a dilated convolution based Wavenet [16] as the generative model. However, results indicate that white noise is superimposed on the output, indicating that the method does not guarantee a backwards compatible quality improvement. In [13], a backward-compatible method was proposed for improving the speech quality of the Opus codec. The method utilized decoded parameters to condition the autoregressive generative models, specifically WaveNet and RNN-based LPCNet (developed by the authors themselves).

A recent breakthrough in generative modeling are generative adversarial networks (GAN) [17]. GANs have achieved tremendous success in the field of computer vision to generate high-resolution diverse images from complex datasets [18], image restoration [19], as well as for speech enhancement [20]. Related to audio coding, a bandwidth extension method based on GANs was proposed in [21]. The method uses a GAN-based generative model to fit the distribution of the MDCT coefficients in the high-frequency bands. In other words, a GAN was used as a technology replacement for the spectral bandwidth replication (SBR) tool as used in the HE-AAC family of audio codecs [22],[4].

The main goal of our work is to improve the sound quality of low-bit rate coded audio without modifications to the existent standard-compliant encoders. Thus, it is important for us to make sure that the proposed generative model does not degrade the default audio quality of the codec. We aim to recover the characteristics that can preserve the sound quality of the original speech and applause content by removing compression artifacts. We specifically focused on these two classes of signals as they are the most challenging for transform codecs.

To the best of our knowledge, GANs have not yet been applied to any coded audio enhancement task, so this is the first contribution to demonstrate an adversarial framework to enhance coded audio. Furthermore, we are not aware of any deep learning based coded applause restoration technique. Unlike previous approaches based on generative models, our method works entirely end-to-end, directly on decoded audio samples. Therefore, no manually-crafted features are extracted and, with that, no explicit assumptions about decoded audio are done (i.e. our method is codec-agnostic). The generator architecture is fully 1D-convolutional, thus providing a fast way to perform forward operations, as we process the full signal with one forward operation through the generator network. This approach contrasts with that of autoregressive models, which cannot be parallelized when computing each time step. Our contribution is based on the speech enhancement GAN (SEGAN) [20]. We chose to enhance AAC coded audio, because AAC (or variants of AAC) is operating at the core of the state-of-the-art audio codecs [22],[4].

Section 2 describes the SEGAN model. Section 3 describes the dataset, model setup, and the training procedure. Finally, the performance of our enhancer is evaluated in Section 4. We then conclude in Section 5.

## 2. SEGAN

The original GAN setup [17] was designed to generate realistic data from statistically independent random noise $\mathbf{z}$. Intuitively, for our task, a generator (G) network should map coded audio $\tilde{\mathbf{x}}$ and $\mathbf{z}$ to its enhanced version $\mathbf{x}^* = G(\tilde{\mathbf{x}}, \mathbf{z})$; and a discriminator (D) network should try to predict the difference between $\mathbf{x}^*$ and original unencoded content $\mathbf{x}$, such that G will receive useful feedback and adapt itself to achieve a realistic enhanced result. Note that generating an enhanced signal $\mathbf{x}^*$ is dependent on the input coded signal $\tilde{\mathbf{x}}$. Conditional GANs [23] take this explicitly into account by training D with both signals as an input (Fig. 1). This enables D to learn the conditional classification task, i.e. whether its input is the original or enhanced signal, based on the given coded signal. This principle was also employed in the SEGAN [20], whose basic aspects (model architecture and loss functions) are described next.

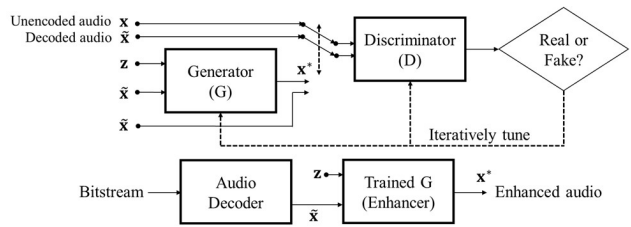

**Fig. 1.** Conditional GAN training setup (top); and deploying the trained G for enhancing decoded audio from a bitstream (bottom).

The encoder of the G network embeds coded audio $\tilde{\mathbf{x}}$ into a deterministic code $\mathbf{c}$ using layers of 1D-convolution (with stride $s$), with Parametric-ReLU (PReLU) nonlinearity in each layer. The latent space is a concatenation of $\mathbf{c}$ and a random vector $\mathbf{z}$ drawn from a prior distribution $\mathcal{N}(\mathbf{0}, \mathbf{I})$. Thus, there is an inherent stochastic component in G that implies a different outcome for every generated prediction. Then, the reconstruction is made in the decoder of G by deconvolving the latent signals back to the time domain using layers of 1D-transposed convolution (with stride $s$) with PReLU in each layer except the last (which has a tanh). Furthermore, the G network features skip connections, routing feature maps from the encoder layers to its homologous decoding layers. Our skip connections are concatenated with decoder feature maps rather than added. Note that this architecture provides a fast way to perform forward operations, as we process the full signal with one forward operation through the G.

The architecture of the D network is nearly the same as the encoder part of G. The only differences are that: (1) it gets two input channels (see, Fig. 1); (2) it uses batch normalization before PReLU nonlinearities; and (3) convolutional layers are followed by two fully connected (FC) layers (with PReLU in each layer) and a final FC layer outputting a single neuron indicating real or fake.

D is a learnable comparative loss function between original and coded audio. Thus, it has the paired input $(\mathbf{x}, \tilde{\mathbf{x}})$ as a real batch and $(\mathbf{x}^*, \tilde{\mathbf{x}})$ as a fake batch. G adapts itself to make the fake batch $(\mathbf{x}^*, \tilde{\mathbf{x}})$ to be classified as real, thus being the adversarial objective. Note that we are using the least-squares GAN [24] form in the adversarial component, so our loss functions for D and G respectively, become

$$\mathcal{L}_D = \frac{1}{2}\mathbb{E}_{\mathbf{x},\tilde{\mathbf{x}} \sim p_{data}(\mathbf{x},\tilde{\mathbf{x}})}[(D(\mathbf{x}, \tilde{\mathbf{x}}) - 1)^2]$$
$$+ \frac{1}{2}\mathbb{E}_{\mathbf{z} \sim p_z(\mathbf{z}), \tilde{\mathbf{x}} \sim p_{data}(\tilde{\mathbf{x}})}[D(\mathbf{x}^*, \tilde{\mathbf{x}})^2], \qquad (1)$$

$$\mathcal{L}_G = \frac{1}{2}\mathbb{E}_{\mathbf{z} \sim p_z(\mathbf{z}), \tilde{\mathbf{x}} \sim p_{data}(\tilde{\mathbf{x}})}[(D(\mathbf{x}^*, \tilde{\mathbf{x}}) - 1)^2]$$
$$+ \lambda \|\mathbf{x}^* - \mathbf{x}\|_1, \qquad (2)$$

where $\mathbf{x} \in \mathbb{R}^T$ is the original unencoded audio, $\mathbf{x}^* \in \mathbb{R}^T$ is the enhanced audio, and $[D(\mathbf{x}, \tilde{\mathbf{x}}), D(\mathbf{x}^*, \tilde{\mathbf{x}})]$ are the discriminator decisions for the real and fake pair. All these signals are vectors of length $T$ samples except for D outputs, which are scalars. The $L_1$-norm regularization weight $\lambda$ (a hyperparameter) helps in stabilizing the initial adversarial training process. It does so by discouraging G from exploring amplitude regions which could cause D to converge to easy-to-discriminate solutions for the fake adversarial cases.

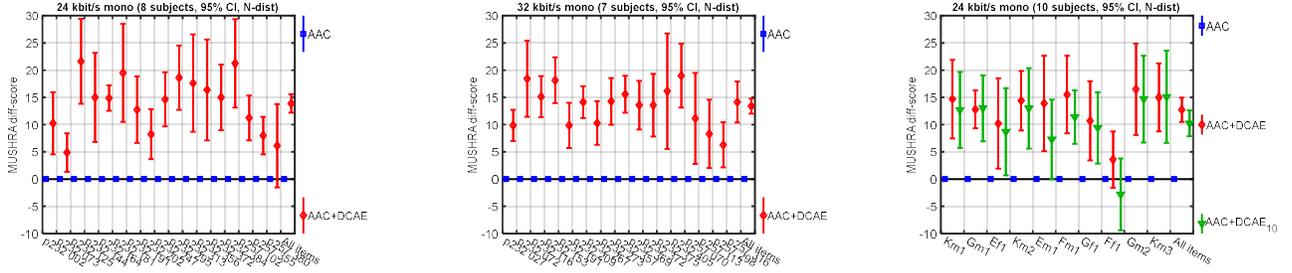

**Fig. 2.** MUSHRA differential results of 24 kbit/s (left) and 32 kbit/s (center) with excerpts from the VCTK speech test set and from an out-of-domain speech set at 24 kbit/s (right). Differential results with DCAE enabled (red) is shown with respect to default AAC without DCAE (blue). Differential results of dimensionality reduced model (green) is shown with respect to default AAC (blue).

## 3. EXPERIMENTAL SETUP

In this section, we describe the experimental setup to evaluate the performance of our coded audio enhancer; this includes the dataset used, model architecture, training setup, and the process for using the trained model to enhance coded audio.

### 3.1. Dataset

The dataset we used for speech [25] consisted of 30 speakers (15 male, 15 female) from the clean Voice Bank [26] dataset (popularly known as the VCTK). The dataset consists of a mix of regional English accents. Of the 30 speakers, 28 (14 male, 14 female) were used to construct the training set, and two were used to build the testing set.

For applause, we analyzed available sound effects CDs referenced in [7] and found that most of the applause signals are contaminated with crowd noise and are not necessarily difficult for coding. Therefore, we created an applause dataset by manually extracting snippets with high perceptual entropy [1] (implying high coding difficulty) from various live music recordings, and from the BBC Sound Effects [27] dataset. In total, the applause training dataset is around 4 hours. Importantly, the test set is totally unseen by (and different from) the training set.

Original speech and applause datasets were recorded at 48 kHz and were down-sampled to 16 kHz. The training set was constructed by encoding (and decoding) the original audio with a tuned AAC encoder operating at 24 and 32 kbit/s mono (at 16 kHz sample rate), with a coding bandwidth of 7.2 and 8 kHz, respectively. Thus, for each unencoded signal we have the corresponding coded version.

### 3.2. Model configuration

To simplify our first modeling approach, we trained two content-dependent enhancer models: one for speech and one for applause.

We used 1D-convolutional layers (without bias terms) in G and D with a stride $s = 2$. The feature maps were incremental in the encoder and decremental in the decoder, having {16, 32, 32, 64, 64, 128, 128, 256, 256, 512, 1024, 512, 256, 256 128, 128, 64, 64, 32, 32, 16, 1} in G (with a total of 22 layers) and {16, 32, 32, 64, 64, 128, 128, 256, 256, 512, 1024} in the convolutional structures of D. Every convolutional layer had a kernel size of 31. We named this model deep "coded audio enhancer" (DCAE). Both D and G were trained for 110 epochs with RMSprop optimizer with a learning rate of $5 \times 10^{-5}$, using a batch size of 64, $\lambda = 100$, and we decayed $\lambda$ at a rate of $1 \times 10^{-5}$ from epoch 100.

During training, we extracted chunks of waveforms with a sliding window of 1.024 s ($T = 16384$ samples at 16 kHz) with 50% overlap. During inference, we slid the window with no overlap through the whole duration of the signal and concatenated the results at the end. All signals processed either in the input of G or D, were pre-emphasized with a 0.95 factor (like in [20]). When we generated data out of G, we de-emphasized it correspondingly.

To evaluate the impact of enhancement quality with a reduced model size, we also trained a model with a reduced dimension. In the smaller model, we reduced the number of layers in G and D, by increasing the stride ($s = 4$). The feature maps in the encoder and decoder being {64, 128, 256, 512, 1024, 512, 256, 128, 64, 1} in G (with a total of 10 layers, thus named it $DCAE_{10}$) and {64, 128, 256, 512, 1024} in D. All other details remained exactly same as DCAE.

For applause enhancement, we initially applied the same models that were developed for speech. However, we were faced with two challenges. First, since the applause dataset was smaller, we found it hard to train the adversarial networks with the bigger DCAE model. Therefore, we proceeded our investigations with $DCAE_{10}$. Second, we found it tricky to balance transient prominence without making applause signals sound dry or artificial. The solution we employed is to start decaying $\lambda$ from epoch 30 (instead of from epoch 100 for speech). The intuition behind this idea is to incorporate a little bit more stochastic behavior in the generator output (than speech). The model was also trained longer, for 130 epochs. Furthermore, since the spectral tilt of applause is different from speech, we disabled the pre- and de-emphasis filter.

## 4. RESULTS

In this section, the performance of the proposed enhancer is evaluated in terms of both subjective and objective quality.

### 4.1. Subjective experiments

In the following, we evaluated the impact of our enhancer on perceptual quality with MUSHRA [28] listening tests. All the listening tests below included a hidden reference, a 3.5 kHz low-pass anchor, and the systems under evaluation. All tests were performed by experienced listeners using headphones.

First, in two separate listening tests we evaluated the quality at 24 and 32 kbit/s without and with DCAE. Seventeen speech excerpts in the two listening tests were randomly chosen from the VCTK test set. As can be seen from the differential MUSHRA scores in Fig. 2 (left and center), with DCAE there is a significant average improvement of 14 MUSHRA points at both 24 and 32 kbit/s. For some excerpts, improvements as high as 22 and 19 MUSHRA points can be observed at 24 and 32 kbit/s, respectively. Overall, in both the listening tests none of the items are significantly degraded.

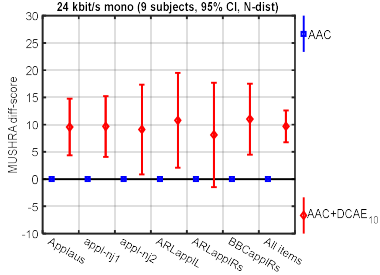

**Fig. 3.** MUSHRA differential results of 24 kbit/s mono applause excerpts with $DCAE_{10}$ enabled (red); with respect to AAC without $DCAE_{10}$ (blue).

Next, we evaluated the performance of the enhancer on excerpts which are outside of the style of the recordings in our dataset, e.g. different languages and accents. For this purpose, we made use of excerpts typically used for audio codec evaluation. Furthermore, in the same listening test we evaluated any impact on audio quality created due to a reduced model dimension ($DCAE_{10}$). Fig. 2 (right) depicts the differential MUSHRA scores. In the item mnemonics, the first letter indicates a language: English, German, Korean or French; the second letter indicates female or male speakers. As can be observed again, all the speech items are improved with the proposed enhancer. Improvements as high as 16 and 15 points are observed with DCAE and $DCAE_{10}$, respectively. Overall, there is a significant average improvement of 13 and 10 points with the DCAE and the $DCAE_{10}$ model, respectively, while none of the items are significantly degraded. These three listening test results demonstrated that our generative network can consistently enhance the quality of coded speech regardless of whether the input comes from a speaker or style seen during training or from an out-of-domain scenario. However, with $DCAE_{10}$ we noted a drop in average audio quality by 3 MUSHRA points.

In the final listening test, we chose applause excerpts and compared the performance at 24 kbit/s without and with the enhancer. We chose only six excerpts in the test because according to our experience, listening continuously to applause excerpts can be fatiguing. Five of the six items chosen in the test included distinct clapping sounds. Only the excerpt ARLapplRs, is noisy and lacks distinct claps. Two of the items (appl-nj1 and appl-nj2) were professionally recorded in a local jazz concert which included cheering in the background. Fig. 3 depicts the differential MUSHRA scores. It can be observed that $DCAE_{10}$ provides a significant improvement for all the excerpts except one. With $DCAE_{10}$, improvements as high as 11 MUSHRA points can be observed. Overall, there is a significant average improvement of 10 points while none of the items are significantly degraded. It is expected that there is less potential to improve a noisy applause excerpt (ARLapplRs); but it is also important to note that our method did not remove so much background noise such that it sounded dry or artificial. Similar results are expected for other signals containing dense transient events (e.g., rain, crackling fire, etc.).

We attribute the improvements with our enhancer due to the following. First, the model performs spectral and temporal noise shaping. Second, due to the generative nature of the model, it can perform spectral gap filling (compare spectrograms in Fig. 4). For applause, we did not observe any effect of noise shaping; rather our model performed a transient-to-noise ratio restoration. Specifically, transients and noise were very slightly (between 0 and 1 dB) amplified and attenuated, respectively. This is in line with the manually-crafted method proposed in [6] followed by transient and noise classification. The advantage of our proposed method is that our method operates entirely end-to-end.

### 4.2. Objective evaluation

We evaluated the quality of the test excerpts from previous listening tests using an objective quality assessment tool. It is known that none of the well-established objective quality tools were designed to evaluate signals synthesized by non-deterministic generative models. In fact, it was shown in [29] that the enhanced quality achieved with a generative decoder was not predicted by the objective tool. We still conducted this evaluation to understand the performance with an objective quality predictor. We computed the mean opinion score - listening quality objective (MOS-LQO) measures predicted by ViSQOLAudio (ViA) [30]. We found ViA to be suitable for our purpose as it was designed to predict the quality of both speech and audio. Like other objective tools, ViA compares the codecs under test with the unencoded originals. The MOS-LQO score ranges from 1–5 (higher score implying better quality). The average MOS-LQO scores are shown in Table 1. We noted that ViA reflected the listening impressions albeit by a small amount.

**Table 1:** Average ViSQOLAudio MOS-LQO scores for conditions and excerpts from the listening tests in Fig. 2 and 3.

| Test set | VCTK | | Out-of-domain | Applause |
|---|---|---|---|---|
| Rate (kbit/s) | 24 | 32 | 24 | 24 |
| AAC | 3.3790 | 3.7037 | 3.7585 | 3.7568 |
| AAC+DCAE | **3.4448** | **3.7508** | **3.8399** | - |
| AAC+$DCAE_{10}$ | - | - | 3.8220 | **3.7643** |

### 5. CONCLUSIONS

We proposed a GAN-based coded audio enhancer for effectively restoring signals contaminated with coding noise. The model has been demonstrated to provide significantly improved perceptual quality for speech and applause signals. Our method directly operates on decoded waveforms, and thus the concept is codec-agnostic. Furthermore, our generative model enables faster processing. An un-optimized PyTorch implementation of our best performing model for speech and applause runs at 5x and 7x real-time, respectively, on a CPU.

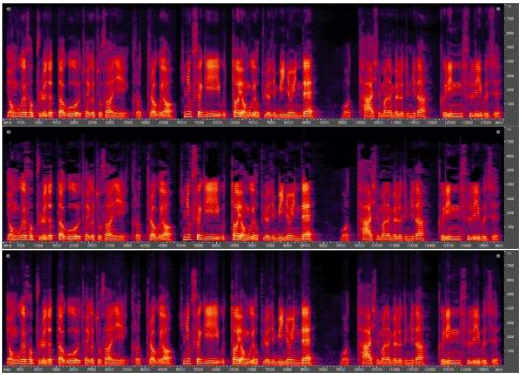

**Fig. 4.** Spectrogram of the original (top), coded (middle) at 24 kbit/s, and corresponding enhanced (bottom) version.